\documentclass[useAMS,usenatbib]{mn2e}
\def\1rx{\hbox{1RXS\,J232953.9+062814}}
\title[Search for brown-dwarf secondaries in CVs II]{Search for brown-dwarf like secondaries in
cataclysmic variables II}
\author[R.E.\ Mennickent, M.P. Diaz and C. Tappert]
{R. E. Mennickent$^1$\thanks{E-mail: rmennick@stars.cfm.udec.cl}
M. P. Diaz$^2$ and C. Tappert$^{1}$\\
$^1$Dpto. de F\'{\i}sica, Facultad de Ciencias F\'{\i}sicas y Matem\'aticas,
Universidad de Concepci\'on, Casilla 160-C, Concepci\'on, Chile\\
$^2$Instituto  Astron\^omico e Geof\'{\i}sico,
Universidade de S\~ao Paulo, Brazil}
\begin{document}

\date{Accepted XXX. Received XXX; in original form XXX}

\pagerange{\pageref{firstpage}--\pageref{lastpage}} \pubyear{2003}

\maketitle

\label{firstpage}

\begin{abstract}
We have examined VTL/ISAAC 1-2.5 $\umu$m spectroscopy
of a sample of short orbital period cataclysmic variables which are
candidates for harboring substellar companions. We provide descriptions
of the infrared  spectrum of \hbox{EI Psc}, \hbox{V834 Cen},
\hbox{WX Cet}, \hbox{VW Hyi}, \hbox{TY PsA} and \hbox{BW Scl}.
Fitting of the IR spectral energy distribution (SED) was
performed by
comparing the observed spectrum with late-type templates.
Absorption features of the secondary star were detected in
\hbox{EI Psc} and \hbox{V834 Cen}, consistent with
dwarf secondaries of spectral type K\,5 $\pm$ 1 and M\,8 $\pm$ 0.5, respectively.
In addition, we report the first
detection of the secondary star in \hbox{VW Hyi}. The SED in this case is well
matched by an  L\,0 $\pm$ 2  type secondary contributing
23 per cent to the overall flux at $\lambda$ = 1.15 $\umu$m.  This is a
surprising result for a system with a relatively high mass transfer rate.
We discuss the implication of our findings on the current
scenarios for cataclysmic variable star evolution.

\end{abstract}

\begin{keywords}
 Stars: individual: \hbox{EI Psc}, \hbox{V834 Cen}, \hbox{WX Cet},
\hbox{VW Hyi}, \hbox{TY PsA}, \hbox{BW Scl}, Stars: binaries: close,
Stars: Cataclysmic Variables, Dwarf novae, fundamental parameters, evolution,
Stars: binaries: general, Stars: low-mass
\end{keywords}

\section{Introduction}

Cataclysmic variable stars (CVs) are semi-detached binaries
consisting of an accreting white dwarf and a
red dwarf donor transferring matter to the
compact object via the inner Lagrangian point.
The orbiting gas interacts with
itself, dissipating energy by viscous
forces and forming a luminous accretion disk around the white dwarf. The spectral
energy distribution of CVs in X-rays and ultraviolet is dominated by
white dwarf and inner disk emission, whereas the accretion disk
contribution is dominant
in the optical and possibly in the near infrared.
In some cases the emission
from the accretion disk in the IR is approximated by a power-law
\citep{b16}. However, in
general we expect a much more complex emission, especially from low-luminosity disks
which may contain extended optically thin regions. On the other hand,
the secondary star might contribute significantly in
the infrared. The determination of the secondary mass in CVs is key to
understand the secular evolution of these objects. Current theories state
that the process of mass transfer becomes linked with the loss of orbital
momentum, so that the binary period becomes shorter while
the hydrogen rich secondary becomes less and
less massive, eventually being eroded by the process, resulting in a kind
of brown dwarf star when the orbital period approaches 80 minutes \citep{b7}.
An alternative scenario considers that most CVs may not yet have
had time to evolve to their theoretical minimum orbital periods.
In this case, for initial configurations with
intermediate secondary masses, thermal-timescale mass transfer
may occur, eventually producing ultrashort-period
systems with a low-mass, hydrogen poor and probably degenerate
secondary  \citep{b61}. One must keep in mind that these secondaries
are formed very differently from field brown dwarfs.
From the above it is evident that relevant
observations for probing current theories of CV evolution should focus on the
determination of the physical parameters of the secondary star for systems
below the orbital period gap.

Observational methods to search for undermassive secondary stars in
cataclysmic variable stars  have been summarized in Paper I.  The empirical evidence available for this class of objects
has been critically examined by  \citet{b62}. They conclude that no
{\it direct} \footnote{Indirect methods include
radial velocity studies,
modeling of the spectral energy distribution and the use of a
superhump period-mass ratio relation.}
evidence for brown-dwarf secondaries in CVs  exists
in the literature. Specifically they show a Keck-II
spectrum of \hbox{LL And} with no evidence for the absorption features
previously claimed by \citet{b8}. However, recently
\citet{b63} reported the detection of a
carbon-deficient brown dwarf-like secondary in \hbox{EF Eri}.
 In this paper we continue the search for brown dwarfs in CVs
initiated by  Mennickent \& Diaz  (2002, Paper I)
by analysing the infrared
spectral energy distribution of a sample of 6 CV candidates for systems at
late evolutionary stages. The objects were selected
for their orbital periods being below the period gap. The sample
includes SU UMa type dwarf novae as well as magnetic CVs in their
photometric low states.

One should
mention that the determination of basic properties of
secondaries in CVs by comparison of their spectra with field stars
templates is intrinsically uncertain. These results are prone to
illumination and heating of the companion photosphere. In addition,
the absorption spectrum may be affected by the filling of some lines with
emission components. On the other hand, a direct
comparison with line strengths or spectral indices specially
designed for classifying late-M and L dwarfs is unreliable
due to the unknown nature of the continuum emission sources.

In the next section we describe the IR spectroscopic observations while in
Section 3 a description of each spectrum is given. A brief discussion of the
observational results is made in Section 4. Conclusions and
future prospects are outlined in Section 5.

  \begin{table*}
 \centering
 \begin{minipage}{90mm}
\caption{Journal of observations. The total integration time, in seconds,
is given for
each spectral band. }
 \begin{tabular}{@{}llccc@{}}
  \hline
   Star &  UT-date &  $J$ & $H$  & $K$  \\
 \hline
\hbox{EI Psc}  &2002-08-10 &3600 &3600 &3600 \\
\hbox{V834 Cen} &2002(06-23(J,K),08-09(H))   & 2400   &2400    & 2400   \\
\hbox{WX Cet} &2002-08-10     &4320    & 4320   &  -   \\
\hbox{VW Hyi} &2002-08-13       &480    & 480   & 480     \\
\hbox{TY PsA} &2002-08-10        & 1600   & 1200   &  800     \\
\hbox{BW Scl} &2002-07-22         & 2400   &2400    &   -     \\
\hline
\end{tabular}
\end{minipage}
\end{table*}

  \begin{table*}
 \centering
 \begin{minipage}{90mm}
\caption{Equivalent widths of absorption lines. }
 \begin{tabular}{@{}ccccc@{}}
  \hline
   Ion &  $\lambda$ ($\mu$) & \multicolumn{3}{c}{$EW(\lambda)$ (\AA)}  \\
       &                    &  \hbox{VW Hyi} & \hbox{V834 Cen} & \hbox{EI Psc} \\
 \hline
\hbox{Na\,{\sc i}} & 1.1404,1.1381 &4.1 &3.3 &-\\
\hbox{K\,{\sc i}}&1.1690 &1.7 &1.3  &-\\
\hbox{K\,{\sc i}}&1.1777,1.1773 & 1.6 &2.0  &-\\
\hbox{K\,{\sc i}}&1.2432 &2.8 &1.2  &-\\
\hbox{K\,{\sc i}}&1.2522 &1.9 &1.6  &-\\
\hbox{K\,{\sc i}}&1.5167,1.5172 &0.7 &-  &-\\
\hbox{Mg\,{\sc i}}&1.5770,1.5753,1.5745 &- &- &3.6 \\
\hbox{$^{12}$CO(8,5)} &1.6620 &- &- &5.5 \\
\hbox{OH} &1.6890 &- & - &4.2 \\
\hbox{Na\,{\sc i}} &2.2062,2.2090 &7.0 &8.0  &6.0\\
\hbox{Ca\,{\sc i}} &2.2614,2.2631,2.2657 & - & - &3.0 \\
\hbox{ $^{12}$CO(2,0)} &2.2935 &- &2.4 &- \\
\hbox{Na\,{\sc i}}& 2.3355,2.3386  &- &- &11.0 \\
\hline
\end{tabular}
\end{minipage}
\end{table*}

\section{Observations and data analysis}

The infrared spectroscopic observations reported in this paper were
obtained at ESO with VLT-Antu using the ISAAC spectrograph in service mode. The data
were taken under clear atmospheric conditions.
An observing log is given in Table 1.
Spectra in the $J, H$ and $K$ bands were obtained
with full width at half maximum ($FWHM$) resolutions ranging from 24 ($J$) to 46 ($K$) angstroms.
Pipeline reduced spectra were used, but
re-calibrated in wavelength by measuring the location
of atmospheric OH emission lines \citep{b52} in the  sky
background.

The telluric absorption features were removed with the aid of
an absorption template available in the ISAAC web page, properly
degraded in resolution to the instrumental resolution of our objects.
The IRAF task ``telluric" was used to find the best scale and shift factors which, when applied to the
normalized telluric template, provided a reasonable correction of the
telluric absorptions in standards and science exposures. This procedure
worked well, except for the regions between 1.35-1.44 microns  and
1.80-1.94 microns, characterized by heavy telluric absorption. These
regions, corresponding to the wavelength limits of the $J, H$ and $K$
spectra, were excluded
from the analysis and are not shown in this paper.
 
A rough flux calibration was performed using observations of the B2IV type
telluric standard \hbox{Hip\,087287}, made with the same instrumental setup
by the ESO operation team as part of the service mode program. Since the
slit losses might be significant for our science objects,  this calibration
is intended to provide a rough correction for the instrumental response only,
leaving the photometric zero point uncertain. The flux of the standard at each
wavelength was determined  using a black body function set to the same
effective temperature of the standard and matching the reported
infrared magnitudes.

Prior to measuring line parameters, the spectra were normalized by fitting
a low-order polynomial function to the continuum.
Then we measured  the equivalent
width, the $FWHM$, the full width at
zero intensity ($FWZI$)
and in cases of fully resolved double peaked emission lines, the full
peak separation $\Delta \lambda$. In these cases we also measured the (continuum normalized)
intensity of the violet ($V$) and red ($R$) emission components as
defined by a double gaussian fit to the emission line.
It should be noted
that the equivalent widths of absorption lines in late type stars
are very hard to determine, since  most lines
are superposed at some level on molecular bands and in many cases
also blended with other atomic lines. We used the same procedure
for each line in different
objects trusting that these widths should be at least internally
consistent.

\subsection{Infrared spectral fitting}

An attempt to quantify the properties of the IR spectra
of
the targets where a secondary
was detected (see next section)
was made
by employing a numerical fitting procedure.
The spectral energy distribution (SED) in the IR was tentatively
parameterized by adding the contribution from a late-type template spectrum
and power-law component. Although the emission of the disk should
differ substantially from a power-law in the IR, we introduced this component
as a first  approximation to the accretion disc continuum.

While the power law is smooth in our wavelength domain and
basically affects the slope of the IR continuum, the detailed shape of
the synthetic continuum in the $J$, $H$ and $K$ bands is strongly dependent on the
stellar template contribution. A nonlinear least squares fitting procedure
was calculated using the following equation:\\

$ S(\lambda) = a \times T(\lambda) + b \times \lambda^{c}$ \hfill(1)\\

\noindent
where $S$ is the observed spectrum, $T$ the red dwarf template spectrum,
$\lambda$ the wavelength in microns
and $\it{a}$,$\it{b}$,$\it{c}$ parameters to be
found. The parameters $a$, $b$ and $c$ were adjusted to minimize the reduced Chi-square
between the observed spectrum and a model fit.
Of course, $\it{a}$ and $\it{b}$ are constrained to the positive domain.
The above procedure has proven to be useful to estimate the
temperatures of secondaries of short orbital period
cataclysmic variables (Paper I).
The data fit ranges were selected carefully to avoid emission
lines and deep telluric bands.
A sequence of template spectral types between M\,3 and L\,7 with a
stepsize of typically
2 subtypes  was taken from \citet{b14} and \citet{b9}. For the case of
 \hbox{EI Psc} (see next section), we used a grid of
 (continuum normalized) $H$-band spectra, provided by \citet{b57}, that
 covers the spectral types between K\,2 and M\,3.
The resolution of the templates was matched with those
of the science spectra
by convolving the data with a Gaussian with the appropriate
$FWHM$. Due to the low rotational velocity of the templates,
it was not necessary to account for rotational broadening
when fitting the spectrum. In practice, the fitting of the
$J$ and $H$ spectral bands provided most of the information
in this method, while the use of the $K$ band was limited to
spectral line analysis due to the lack of adequate templates
at the required instrumental resolution.

 \begin{table*}
 \centering
 \begin{minipage}{90mm}
\caption{Spectroscopic data for the main emission lines. $V,R$
are the continuum
normalized intensities of the violet and red emission components
of the double emission
lines and $\Delta \lambda$ is the peak separation. }
 \begin{tabular}{@{}llrcccc@{}}
  \hline
   Star &  Line &  $EW$ & $\Delta \lambda$  & $FWHM$ & $FWZI$ & $V,R$ \\
        &       &  (\AA) &\multicolumn{1}{c}{(km s$^{-1}$)} &\multicolumn{1}{c}{(km s$^{-1}$)}  &\multicolumn{1}{c}{(km s$^{-1}$)}         & \\
 \hline
\hbox{EI Psc}   & Pa\,$\beta$ &-20  &1085 &1800 &3205 &1.28,1.22 \\
 \hbox{EI Psc}   & Br\,$\gamma$&-26 &- & -&2860 & - \\
\hbox{WX Cet} & Pa\,$\beta$ & -93 &-  &1920 &3745 &- \\
\hbox{VW Hyi} & Pa\,$\beta$ & -22 &940 &1440 &3460 &1.36,1.25 \\
\hbox{VW Hyi} & Br\,$\gamma$& -16 &1010 &-  &3765 &1.18,1.10 \\
\hbox{TY PsA} & Pa\,$\beta$ & -40 &-  &2130 &3440 & - \\
\hbox{TY PsA} & Br\,$\gamma$& -34&-  &1925&2730 &- \\
\hbox{BW Scl} & Pa\,$\beta$ & -100 &1285 &2290 &3090&2.24,2.12 \\
\hline
\end{tabular}
\end{minipage}
\end{table*}

\section{Results}

The infrared spectra for all program stars
are shown in Figs.\,1$-$3. Spectroscopic parameters
are given in Tables 2 and 3.

\subsection{EI Psc}

This recently discovered SU UMa subtype
dwarf nova \citep{b100}, also named RXS\,J232953.9+062814, has a unique short
orbital period of 64 minutes, shorter than any other hydrogen-rich
cataclysmic variable. At quiescence, the spectrum shows
the absorption features of a type K\,4 $\pm$ 2 secondary star \citep{b53}.
Both
the radial velocity study and the observed
period excess indicate a secondary mass around 0.12 $M\odot$
\citep{b53,b54}.
The secondary star is much hotter than main sequence stars of similar
mass, but it is well matched by helium-enriched models, indicating that it
evolved from a more massive progenitor \citep{b53,b54,b55}.
According to \citet{b53}, the secondary star contributes
50 $\pm$ 20 per cent of the light near 5500 \AA.

Our observations reveal weak Paschen\,$\beta$ and
Brackett\,$\gamma$ emission, the former being clearly double
with the violet component stronger than the red one.
The H-band and K-band spectra show absorption features
probably arising from the secondary star. This view is
supported by the comparison with
a K5\,V type template \citep{b57}
rectified to the resolution of our H-band spectra (Fig.\,4).
We identify several
absorption features like \hbox{Mg\,{\sc i}} 1.5760, 1.7113 $\umu$m,
\hbox{Si\,{\sc i}} 1.5964, 1.6685 $\umu$m, \hbox{$^{12}$CO\,{\sc i}} 1.619, 1.662 $\umu$m,
\hbox{Al\,{\sc i}} 1.6742 $\umu$m and \hbox{OH\,{\sc i}} 1.689 $\umu$m.

We performed the fitting of the function
described by Eq.\,(1) in the range
1.52--1.78 $\umu$m, using the
K2--M3 type templates and the continuum
normalized spectrum observed at the $H$-band, forcing $c$ = 0.
This was needed since only continuum-normalized templates were available
for these spectral types. We
obtained  the best fit with a K\,5 $\pm$ 1 type secondary
contributing 50 $\pm$ 10 per cent  to the total flux in the $H$-band.
This number was particularly sensitive to the spectral range considered
in the fit. Including only the absorption lines and excluding the continuum
at the end of the spectrum, and also the "bump" between
1.60$-$1.61 $\umu$m, the estimated secondary star contribution increases to
67 $\pm$ 6 per cent.
The fit is severely degradated for
earlier or later type templates. In the above estimate we have assumed that
the relative disc contribution is nearly
constant through the spectral range considered.
Our result of a K\,5 type secondary is in agreement with the
spectral type reported by \citet{b53}.

\citet{b59} found that the veiling-independent indicator:    \\

$r = \log[\frac{EW(^{12}CO(2,0))}{EW(Na)+EW(Ca)}$] \hfill(2)    \\

\noindent
is a strong luminosity indicator among K-M stars,
independent of effective temperature. The almost null visibility of
$^{12}$CO(2,0) features in our spectra along with the detectable presence of
the sodium doublet and the calcium triplet, point to a very low $r$ value and therefore
a dwarf secondary, as expected in a short orbital period CV.

We have tried to establish the temperature of the secondary
star using the veiling-independent temperature/luminosity
discriminator: \\

$(\frac{EW(OH\, 1.6904\, \umu m)}{EW(Mg\, 1.5765\, \umu m)},\frac{EW(CO\, 1.6610\, \umu m
+ CO\, 1.6187\, \umu m)}{EW(Mg\, 1.5765\, \umu m)})$\hfill(3)  \\

\noindent
as proposed by \citet{b59}. We find  for this discriminator
(1.2 $\pm$ 0.2, 1.5 $\pm$ 0.2). Using Eq.\,(5) from \citet{b57}
yields an effective
temperature,
$T_{\rm eff}$ = 1585 $\pm$ 745 K,
which is incompatible with
a K-type star.
A close inspection of Fig.\,4 reveals that the discrepancy
arises from the anomalously large \hbox{OH} line, which in K-type
stars is normally weaker. A departure of normal conditions
of the irradiated secondary could provide an explanation for the  discrepancy.

\subsection{V834 Cen}

In this paper we present the first observations of the IR spectrum of
this short-period AM Her system. \hbox{V834 Cen} is the only magnetic
CV in our sample. The object has been relatively well studied
in X-ray, UV and optical wavelengths. The
orbital period is 101.5 min \citep{b37} and
the field strength derived from cyclotron and Zeeman
features is B $\approx$ 23 MG \citep{b38}.
According
to the VSNET\footnote{VSNET: Variable Star Network
(http://vsnet.kusastro.kyoto-u.ac.jp/vsnet/index.html)}
records, our observations were obtained during the slow
recovery from a low state. This is consistent with the lack of strong emission
features in the observed spectra.

In principle, polars in
low states are ideal scenarios to detect brown dwarf-like secondaries
since the light from these systems is not "contaminated" by the
accretion flux. Our observations of \hbox{V834 Cen}
reveal an IR flux distribution characterized by a
steep blue continuum and by the presence of
absorption features of the secondary star.
We observe
the \hbox{K\,{\sc i}} doublets at
1.169-1.177 $\umu$m and 1.244-1.253 $\umu$m,
the \hbox{Na\,{\sc i}} lines at 1.141, and 2.206-2.209 $\umu$m
and possibly the \hbox{Al\,{\sc i}}
doublet at 1.313 and 1.315 $\umu$m.
The $^{12}$CO absorption line at 2.2935 is also visible along with the
$^{12}$CO + \hbox{Na\,{\sc i}} complex at 2.3227,2.3355 and 2.3386
$\umu$m.
On the other hand, the \hbox{H\,{\sc i}} lines seem to be filled by emission,
and in Brackett\,$\gamma$ a very narrow and weak emission
($FWZI$ = 920 km/s, $EW$ = -1 \AA) is observed over
the continuum. 
No traces of Zeeman
splitting of the \hbox{H\,{\sc i}} lines are observed, as
occurs in the optical spectra of the AM Her star
\hbox{EF Eri} in its low state \citep{b70,b71}. Such behaviour
in \hbox{EF Eri} suggests that
the accretion flux has not completely vanished and is still
contributing more to the IR continuum than the
white dwarf.

We observe a "bump" in the spectrum starting at $\lambda \sim$ 1.29
$\umu$m (Fig.\,1). This feature could be explained as a cyclotron bump, as those
observed during a faint state in the infrared spectrum of
\hbox{AM Her} \citep{b577}. Unfortunately,
our spectral coverage is inadequate to test this assumption.

For \hbox{V834 Cen}, the luminosity index defined in Eq\,(2),
$-$0.5 $\pm$ 0.1, is compatible with a dwarf type secondary star.
The best spectral fitting for the $J$-band spectrum
was obtained with a M\,8 secondary contributing 30 per cent to the overall
flux at $\lambda$ = 1.24 $\umu$m, and a steep
continuum characterized
by an exponential index $c$ = $-$6.9 $\pm$ 0.1.
This fit is significantly  better than those obtained even with M\,7 and
M\,9.5 type templates. Our secondary spectral type is later than those
estimated by \citet{b39}, viz.\ dM\,5, based on the detection of TiO
bands in the optical spectrum at low state,  and it is at
the cooler limit of the M\,5$-$M\,8 range derived by \citet{b40} from the study of
multi-wavelength photometry. It is also later than the dM\,6.5 classification
provided by \citet{b41} from spectroscopic modeling. However,
these authors did not use templates
with spectral types later than dM\,6.5.
Our best fit is shown along with the object spectrum in  Fig.\,5.
From the arguments above
we suspect that previous identifications of the spectral type of the secondary star in
\hbox{V834 Cen} may suffer from large uncertainties and that the
value of M\,8 derived in this paper   possibly
represents a more robust determination. The continuum slope given
above
implies a very strong dependence on $\lambda$. Just for comparison,
a black body with temperature in the range of 15\,000$-$50\,000 K
should contribute with a continuum with a spectral index $c$
between $-$3.5 and $-$4 in the spectral range considered. We conclude
that there is a bright hot source contributing to the
IR flux that may be related to the weak mass accretion during low states.

\subsection{WX Cet}

This dwarf nova was originally proposed as a WZ Sge-like system,
due to scarce and large amplitude outbursts \citep{b48}.
However, observations during superoutbursts revealed
a rather mild photometric period excess,
supporting the hypothesis that WX Cet is a fairly normal large-amplitude
SU UMa-type dwarf nova, rather than a WZ Sge-type dwarf nova \citep{b45,b47}.
Recent time-resolved spectroscopy
suggests that \hbox{WX Cet} possesses a more
massive secondary and
represents an earlier stage of mass transfer
in a cataclysmic binary than \hbox{WZ Sge} \citep{b46}.
However, the secondary star mass derived by these authors,
viz\,  $M_{2}$ = 0.047 $\pm$ 0.013 $M_{\odot}$,
is still below the accepted value for the hydrogen-burning
minimum mass $\sim$ 0.07
$M_{\odot}$ \citep{b105}.
This is consistent with the recent
result of the multi-component fitting to the
SED by \citet{b49}, who obtained a secondary star temperature
of 850 $\pm$ 150 K, which is, surprisingly,
compatible with a 5 Gyr old brown-dwarf of
0.05 $M_{\odot}$ \citep{b105}.
Evidently, the possibility of a period bouncer is still
open and the star deserves further study, especially in the
infrared.

Our $J$-$H$ band spectra show no sign of the secondary star,  but do
show broad Paschen $\beta$ emission and very weak emission in high order
Brackett lines. Our Paschen $\beta$ emission profile does not reproduce the
strong asymmetry observed in the same line by \citet{b49},
suggesting a variability reminiscent of that
observed in the Balmer emission lines. The fact that no secondary star features were detected
in our IR spectra may suggest that the secondary is particularly
cool and is being
outshined
by the disk emission. Further IR spectroscopy with improved
$S/N$ is encouraged in order to better clarify this point.

\subsection{VW Hyi}

Comparatively few spectroscopic studies exist of this
rather bright southern SU UMa star.  A reddening-based
distance of 40 pc is given by \citet{b101}.
The orbital period was established as 107 min \citep{b51}.
The extensive photometric record existing in the VSNET archives
supports the dwarf nova classification and
evidences the rather short recurrence time of $\simeq$ 27 days
and outburst amplitude of about 5 magnitude.
According to these records, our observations were taken at quiescence,
17 days before maximum.
\hbox{VW Hyi} is one of the few CVs where Doppler tomography
clearly shows the contributions of the different line emitting sources
in the system: accretion disc, gas stream, hot spot and possibly
secondary star
\citep{b50,b64}. In particular, the above study indicates that
the secondary star could significantly contribute to the overall
H$\alpha$ emission during quiescence.

Our $J$-$H$-$K$ band spectra clearly reveal the presence of the absorption
features of the secondary star. We observe
the \hbox{K\,{\sc i}} doublets at
1.169-1.177 $\umu$m and 1.244-1.253 $\umu$m
and the \hbox{Na\,{\sc i}} lines at 1.141, and 2.206-2.209 $\umu$m.
Paschen $\beta$ and Brackett\,$\gamma$ appears like asymmetric double
emissions with the violet peak stronger than the red one.
Strong \hbox{H\,{\sc i}} double emissions also appear in the
IR spectra shown by \citet{b49}, but they are symmetric, and no
evidence of the secondary star is visible in her (lower $S/N$)
spectra.

The best spectral fitting for the $J$ band spectrum
was obtained with an L\,0 type
secondary contributing 23 per cent to the overall
flux at $\lambda$ = 1.15 $\umu$m, and a continuum characterized
by an exponential index $c$ = -0.38 $\pm$ 0.02.
The fit is also good for spectral types M\,8 up to L\,2,
but degradates significantly  outside this range (Fig.\,6).
If we assume $J$ = 12.56 from the 2MASS photometry \citep{b106}, we
get for the secondary star $m_{\rm J}$ = 14.15, and assuming a
distance of 40 pc we obtain $M_{\rm J}$ =  11.14, which is compatible
with a secondary star below the
hydrogen-burning minimum mass limit but with an age less than
0.5 Gyr \citep{b105}.
A late type secondary  for \hbox{VW Hyi}
is a surprising result, in apparent conflict with the
0.11 $M_{\odot}$ secondary found by  \citet{b51}.
However, these authors used a dynamical
model based on the radial velocity amplitude of the
emission lines, which has proven to be, at least for some
systems, an highly unreliable indicator. In fact, Schoembs \& Vogt (1981)
determined a semi-amplitude of the radial velocities $K_{1}$ = 78 $\pm$ 14
km s$^{-1}$, while Tappert et al. (2003) obtained a much
lower value of 38 $\pm$ 9 km s$^{-1}$.
We will discuss the implications of this finding in the next
section.

\subsection{TY PsA}

The quiescent optical spectrum of the SU UMa-type dwarf
nova \hbox{TY PsA} is characterized by broad, double-peaked emission lines \citep{b44}.
A periodic modulation observed in the radial velocities during quiescence
suggests an orbital period of 0.08414 d \citep{b44}.
The absence of radial velocities of the broad
shallow absorption lines during superoutburst
could be explained if \hbox{TY PsA} has
an extreme mass ratio and an unusually low-mass secondary star \citep{b44}.
However, this view apparently conflicts with the  rather large observed
difference between orbital and superhump period \citep{b45}.
To our knowledge, our spectra are the first to reveal the
infrared emission of this CV; showing no absorption
features of the secondary star,
but strong emission in Paschen\,$\beta$ and Brackett\,$\gamma$
and weaker emission in higher order Brackett lines. The central reversal
in the emission line, if present, is quite shallow,
indicating a moderate systemic inclination.

\subsection{BW Scl}

This variable star, also named
\hbox{HE 2350-3908} and \hbox{RX J2353.0-3852},
was discovered independently
in the Hamburg/ESO and the ROSAT surveys
\citep[Augusteijn \& Wisotzki 1997,][]{b43}.
It has one of the shortest orbital periods (78 min)
among CVs with normal hydrogen-rich secondaries.
The optical spectrum and the orbital light curve of \hbox{BW Scl}
is very similar to \hbox{WZ Sge}, suggesting that the source is
a dwarf nova type cataclysmic variable with a very low mass-transfer rate,
and a long recurrence time \citep{b42}. This view has not yet been
tested, partly due to the
lack of an adequate long-term photometric record. However,
our $J$-$H$ band spectra -- the first obtained in this spectral region --
seem to corroborate the dwarf nova classification;
they show a broad double-peaked Paschen\,$\beta$ emission line
and weaker broad emission in the Bracket
series,
typical credentials of dwarf novae. No sign of the secondary
star is visible in our spectra.  The central absorption
observed in Paschen\,$\beta$ suggests a moderate inclination for the
system.  The star shows the strongest Paschen\,$\beta$
emission in our sample, with an
equivalent width comparable to those
shown by the \hbox{WZ Sge} type star candidate
\hbox{1RXS J105010.3-140431} \citep{b3}. The rather
large Pa$\beta$ emission
is compatible with an origin in a
low mass accretion rate disc.

\section{Discussion}

In the above section we have reported
evidence for M\,8 and L\,0 type
secondaries in \hbox{V834 Cen} and \hbox{VW Hyi}, respectively,
and for a K\,5 type secondary in \hbox{EI Psc}.  While
\hbox{EI Psc} has already
been discussed in the context of CV evolution, being interpreted as
the result of mass transfer on a
thermal timescale in a system with a secondary initially
more massive that the white dwarf \citep{b53}, the other cases
are, at first glance,
new candidates for period bouncers. As an illustration, a CV
with orbital period around 105 minutes
should posses a secondary star of spectral type around M\,3
in the upper branch of the CV evolutionary track
\citep{b7}. However,
\hbox{VW Hyi} appears like a bizarre object,
since this star does not show the typical
credentials of a star which has bounced-off from the
orbital period minimum,
viz.\, a low mass transfer rate evidenced in a
long recurrence time and large amplitude outbursts.
For this reason,
the star has never been included in the list of  WZ Sge
type candidates or related systems \citep{b102}.

To our knowledge, \hbox{VW Hyi} is the first
CV showing simultaneously evidence for a late type secondary
and relatively high mass transfer rate.
Based on the evidence, it is hard to conclude
that the star is the result of the
evolution of a shorter period system.
A period bouncer with this orbital period would have
 $\dot{M}$ $\sim$ 4 $\times$ 10$^{-12}$   $M_{\odot}$ yr$^{-1}$
\citep{b7}, while the recurrence time  of 27 days
yields  $\dot{M}$ $\sim$ 5 $\times$ 10$^{-11}$ $M_{\odot}$ yr$^{-1}$, using equation
36 by \citet{b104}.
However, accordingly to current theories,
there is a $\dot{M}$ spike at the onset of mass transfer
(e.g. Howell, Nelson \& Rappaport 2001 and references therein).
Might  \hbox{VW Hyi} be a recently
formed CV with an initial
low mass secondary and transient high $\dot{M}$?
Using the Patterson (2001)
empirical relationship between the period excess
(0.033 for \hbox{VW Hyi})
and the mass ratio we find $q$ = 0.15 $\pm$ 0.01.
If the secondary is a L\,0 type star, its mass should be
around 0.06 $M_{\odot}$ and the mass of the white dwarf
should be 0.4  $M_{\odot}$, which is not an unrealistic
value.

On the other hand, the mass transfer rate  ($\dot{M}$) in \hbox{V834 Cen}
is hard to quantify due to the lack of recurrent outbursts in this
(diskless) polar.  However, the secondary star temperature
inferred
in this work, $T_{2} \approx$ 2200 K,
conflicts with the theoretical prediction for a period bouncer with
an orbital period around 100 min, viz.\, $\approx$ 700 K \citep{b7};
the same
occurs for  \hbox{VY Aqr}, as found in Paper I. Is it possible that
irradiation effects can account for these differences?
If the case of \hbox{EF Eri} \citep{b63} turns out to be representative
for most short period CVs, differences up to 700 K could exist
on the irradiated surfaces of the secondary stars. This
raises the caveat that secondary
temperature determinations based on single IR spectra,
with exposure times a significant fraction of the binary
orbital period
are in fact only upper limits for the
photospheric temperature.
This could be the cases for \hbox{VY Aqr}
and \hbox{V834 Cen}.



\section{Conclusions}

\begin{itemize}
\item We have found evidence for K\,5, M\,8 and L\,0
type secondaries in
\hbox{EI Psc}, \hbox{V\,834 Cen} and \hbox{VW Hyi}, respectively.
This may suggest that \hbox{V\,834 Cen} and \hbox{VW Hyi}
have probably
passed beyond the orbital period minimum. However, the
case of \hbox{VW Hyi} is peculiar, in the sense that
the system also shows a relatively high $\dot{M}$. We prefer to
interpret
this result for \hbox{VW Hyi} in terms of a recently formed CV with a brown-dwarf like
secondary.

\item For some objects, namely \hbox{WX Cet},
\hbox{TY PsA} and \hbox{BW Scl},
we found no significant improvement of
the spectral fitting by adding a stellar atmosphere template.
This may indicate that
even for such low luminosity systems the accretion disk spectrum
dominates the flux in the IR.

\item In this paper and Paper I we find that the
fitting of the infrared spectral distribution
using stellar templates is an useful diagnostic tool
for the secondaries in ultra-short orbital period CVs.
However, there exists evidence that due to irradiation, only
upper limits for the secondary star temperature can be derived
with the current instrumentation, and that exact measurements
of $T_{2}$ should wait for even larger telescopes with the
ability of time-resolved infrared spectroscopy of faint CVs.

\end{itemize}

\section*{Acknowledgments}
 
We acknowledge an anonymous referee for reading
a first version of this manuscript and giving
valuable hints about how to improve its presentation.
We thank Sandy Leggett,
who provided some of the IR digital
spectra of low mass red objects used in our SED models.
REM acknowledges support by Grant
Fondecyt \#1030707 and DI UdeC \#202.011.030-1.0.
MPD thanks the CNPq support under grant \#301029.

Figure  captions

Fig.1 $J$-band spectra. They are normalized to the mean flux at 1.27$-$1.29 $\umu$m,
  and displaced by multiples of 1.5 for better
  visualization. The spectrum of a B-type standard is also shown indicating the position of the
  telluric absorption bands.
  Vertical lines show the rest wavelength of H lines and some spectral features commonly observed
  in late
  M stars.

  Fig. 2 Same as Fig.\,1 for $H$-band spectra. The spectra are normalized
  to the mean flux at 1.54$-$1.56 $\umu$m,  and shifted by
  multiples of 1.
 
Fig.3   Same as Fig.\,1 for $K$-band spectra. The spectra are normalized
  to the mean flux at 2.09$-$2.11 $\umu$m, and shifted by
  multiples of 0.5. The feature observed around 2.186 $\umu$m in the spectrum of
  EI Psc is probably an artifact.

Fig. 4  The continuum normalized $H$-band spectrum of EI Psc
  (thin line) along with the spectrum of the K5\,V-type star \hbox{HR\,8085} (thick line). Features of
  the secondary star are clearly visible in the spectrum of EI Psc.

Fig. 5  The $J$-band spectrum of \hbox{V\,834 Cen} and the best
  composite SED fit (thick line). The individual M\,8 type
  template spectrum
  and power law continuum are also shown.

  Fig. 6 The $J$-band spectrum of \hbox{VW Hyi}, normalized to
  the flux at $\lambda$ = 1.15 $\umu$,
  and the best
  composite SED fit (thick line), for different secondary star
  templates. The best fit is obtained with a
  template spectrum of spectral type L\,0.

\bsp

\label{lastpage}

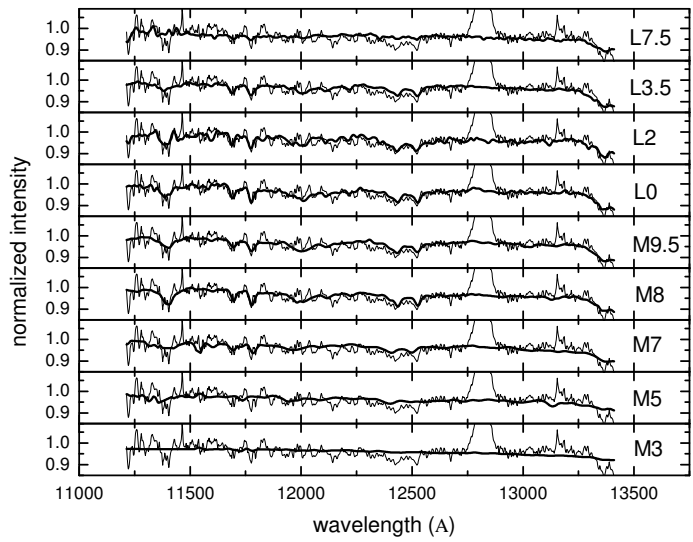
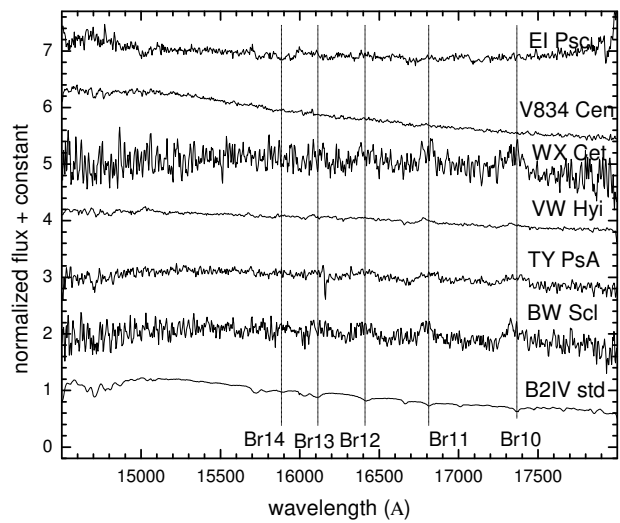
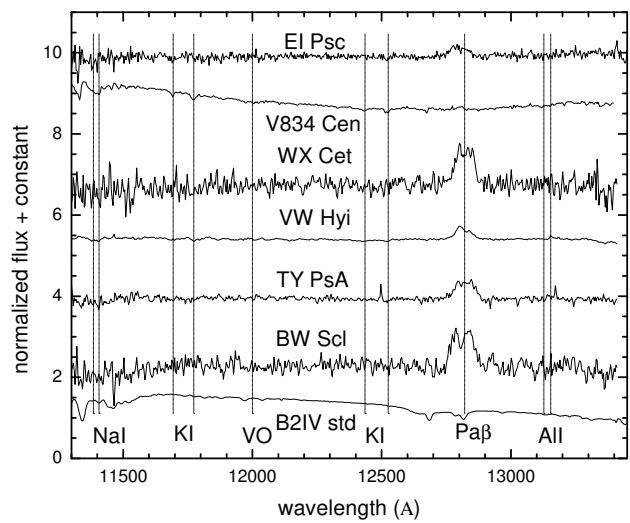
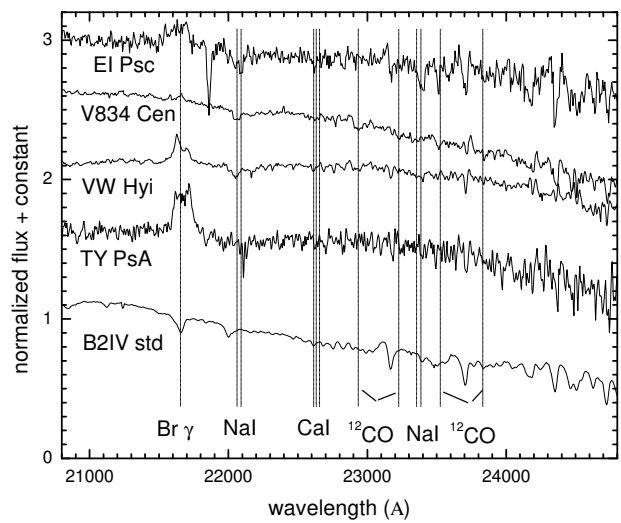
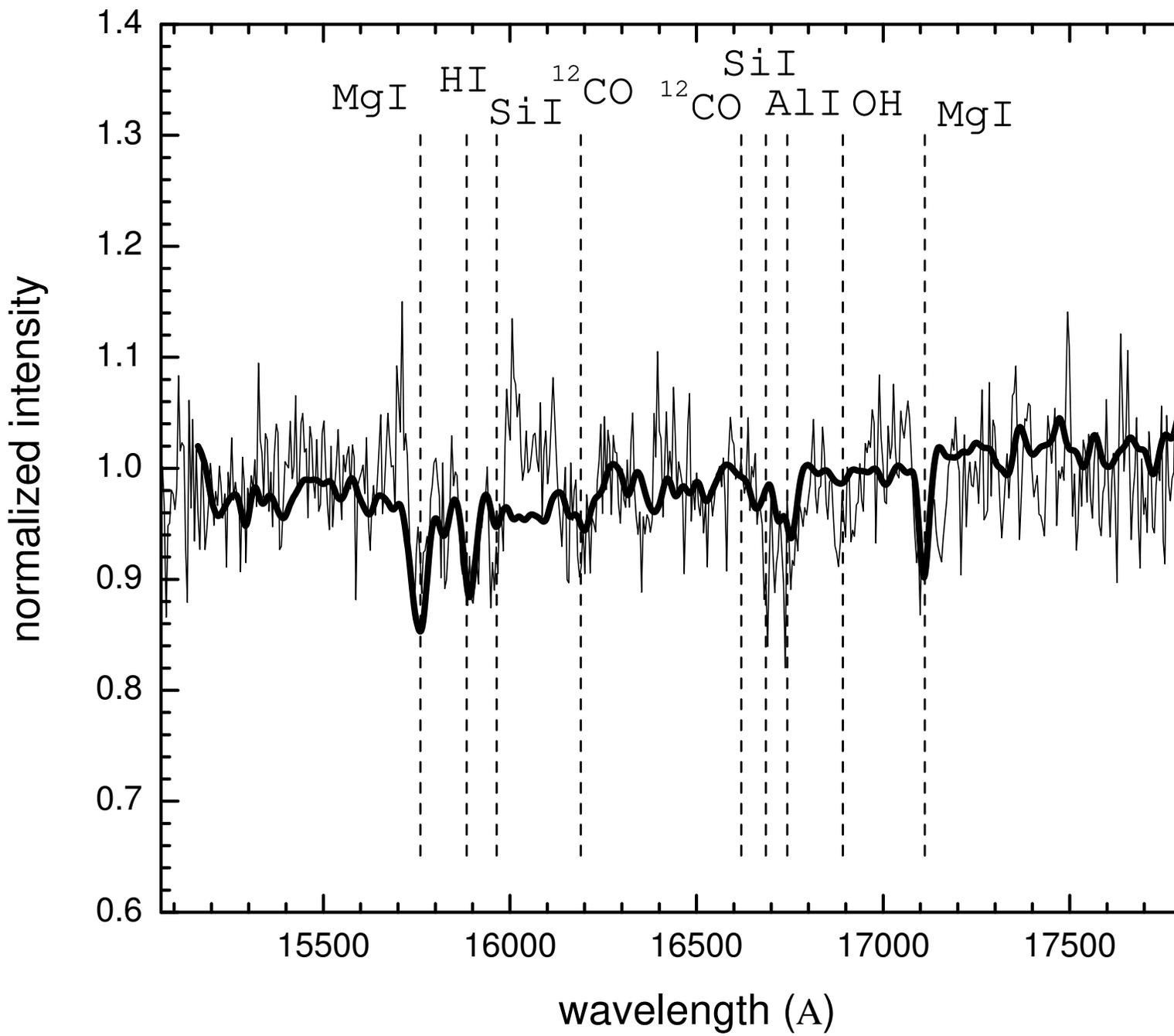
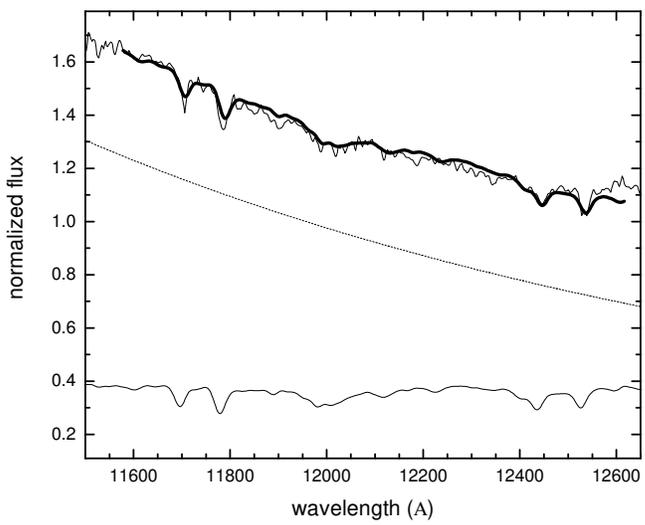
\end{document}